\documentclass[conference]{IEEEtran}
\usepackage{cite}

\ifCLASSINFOpdf
\else
\fi
\usepackage{amsmath}
\usepackage{amssymb}
\usepackage{amsfonts}
\usepackage{amsthm}
\usepackage{graphicx}
\usepackage{bm}
\usepackage{multirow}	
\usepackage{listings}
\usepackage{cases}
\usepackage{array}
\usepackage{booktabs}
\usepackage{algorithm}
\usepackage{algpseudocode}
\usepackage{color}
\newtheorem*{remark}{Remark}
\usepackage{hyperref}
\hypersetup{
     colorlinks   = true,
     linkcolor    = blue,
     citecolor    = red
}
\usepackage[caption=false, font=footnotesize]{subfig}
\usepackage{authblk}

\newcommand{\RN}[1]{%
  \textup{\uppercase\expandafter{\romannumeral#1}}%
}

\makeatletter
\def\bstctlcite{\@ifnextchar[{\@bstctlcite}{\@bstctlcite[@auxout]}}
\def\@bstctlcite[#1]#2{\@bsphack
  \@for\@citeb:=#2\do{%
    \edef\@citeb{\expandafter\@firstofone\@citeb}%
    \if@filesw\immediate\write\csname #1\endcsname{\string\citation{\@citeb}}\fi}%
  \@esphack}
\makeatother

\DeclareMathOperator{\Cov}{Cov}
\DeclareMathOperator{\Var}{Var}
\renewcommand{\d}[1]{\ensuremath{\operatorname{d}\!{#1}}}
\newcommand{\E}{{\mathrm{E}}}

\makeatletter
\newcommand*{\transpose}{%
  {\mathpalette\@transpose{}}%
}
\newcommand*{\@transpose}[2]{%
  \raisebox{\depth}{$\m@th#1\intercal$}%
}
\makeatother

\begin{document}

%

\bstctlcite{IEEEexample:BSTcontrol}

\title{Uncertainty Quantification in Stochastic Economic Dispatch using Gaussian Process Emulation\vspace{-2ex}}

\author[1]{Zhixiong Hu}
\author[2]{Yijun Xu}
\author[3]{Mert Korkali} 
\author[4]{Xiao Chen}
\author[2]{Lamine Mili}
\author[4]{Charles H. Tong}

\affil[1]{Department of Statistics, University of California-Santa Cruz, Santa Cruz, CA 95064 USA\\ e-mail: \scriptsize{\textsf{zhu95@ucsc.edu}}}
\affil[2]{Department of Electrical and Computer Engineering, Virginia Tech, 
Northern Virginia Center, Falls Church, VA 22043 USA\\ e-mail: \scriptsize{\textsf{\{yijunxu,lmili\}@vt.edu}}}
\affil[3]{Computational Engineering Division, Lawrence Livermore National Laboratory, Livermore,
CA 94550 USA\\ e-mail: \scriptsize{\textsf{korkali1@llnl.gov}}}
\affil[4]{Center for Applied Scientific Computing, Lawrence Livermore National Laboratory, Livermore,
CA 94550 USA\\ e-mail: \scriptsize{\textsf{\{chen73,tong10\}@llnl.gov}}}

%
%
%


\maketitle

\begin{abstract}
The increasing penetration of renewable energy resources in power systems, represented as random processes, converts the traditional deterministic economic dispatch problem into a stochastic one. To solve this stochastic economic dispatch, the conventional Monte Carlo method is prohibitively time consuming for medium- and large-scale power systems. To overcome this problem, we propose in this paper a novel Gaussian-process-emulator-based approach to quantify the uncertainty  in  the  stochastic economic dispatch considering  wind  power penetration. Based on the dimension-reduction results obtained by the Karhunen-Lo\`eve expansion, a Gaussian-process emulator is constructed. This surrogate allows us to evaluate the economic dispatch solver at sampled values with a negligible computational cost while maintaining a desirable accuracy. Simulation results conducted on the IEEE 118-bus system reveal that the proposed method has an excellent performance as compared to the traditional Monte Carlo method.
\end{abstract}


%
\IEEEpeerreviewmaketitle
\section{Introduction}
\IEEEPARstart{P}{ower} systems are inherently stochastic. Sources of stochasticity include time-varying loads, renewable energy intermittencies, and random outages of generating units, lines, and transformers, to cite a few. These stochasticities translate into uncertainties in the power system models. To address this problem, research activities have focused on uncertainty quantification in power system planning, monitoring, and control~\cite{safta2016efficient, xu2018propagating,xu2019probabilistic, xu2019response, xu2018maximum,  sheng2018applying}. Among them, the topic of stochastic economic dispatch (SED) has recently attracted considerable academic attention due to the increasing penetration of renewable energy resources. 

To account for these uncertainties, some researchers propose to adopt a scenario-based optimization approach. However, this approach only considers a finite set of sampling realizations, which is obviously an oversimplification of the numerous cases that may occur in reality~\cite{ruiz2009uncertainty,takriti1996stochastic}. By contrast, other researchers propose to make use of uncertainty quantification techniques via Monte Carlo sampling. However, all the traditional Monte Carlo methods are prohibitively time consuming when accurate estimation of uncertain model outputs are needed.  This problem calls for the development of new computationally efficient and accurate uncertainty modeling techniques for power system applications \cite{safta2016efficient,li2018compressive}. 


In this paper, we develop a new SED method based on a Gaussian process emulator (GPE) for power systems to which are connected wind power generation. The GPE allows us to evaluate, with a negligible computational cost, the SED solver at sampled values through a nonparametric reduced-order representation \cite{rasmussen2006gaussian}. To further improve the computational efficiency in the construction of the surrogate models, a model reduction is achieved via the application of the Karhunen-Lo\`eve expansion (KLE) to real-world  data, collected from real-world wind farms~\cite{ghanem2003stochastic}. 
The simulation results conducted on a modified IEEE 118-bus system reveal that the proposed method can greatly improve the computational efficiency of the SED as compared to the traditional Monte Carlo method while maintaining a desirable estimation accuracy.

%


\section{Problem Formulation}
Traditionally, under some physical and economic constraints, the economic dispatch in power systems is known as a deterministic optimization problem. This problem aims to identify an optimal set of power outputs of a fixed set of online thermal generating units that yields a minimum cost, denoted by $Q( \mathbf{g})$. The cost $Q$ is generally thought to be nonrandom since the traditional thermal generating units $\mathbf{u}$ can be optimized and set equal to some deterministic optimal values.  

However, in the face of the increasing penetration of renewable energy resources, the abovementioned statement cannot hold true. Due to the intrinsic randomness of the renewable generation, represented (using random fields) as functions of a vector of random variables, $\bm{\omega}$, denoted by $\mathbf{p}(\bm{\omega})$, the deterministic economic problem for finding $
 Q( \mathbf{g})=\underset{\mathbf{g}}{\arg \min }
  \{f(\mathbf{g})\} $ 
  is extended to an SED problem described by
  \begin{equation}
      Q( \mathbf{g,\mathbf{p}(\bm{\omega})})=\underset{\mathbf{g}}{\arg \min}
  \{f(\mathbf{g,\mathbf{p}(\bm{\omega})})\}.
  \end{equation}
Here, $f$ represents the objective function. For this problem, the randomness brought by $\mathbf{p}(\bm{\omega})$ will lead to different optimized values of $\mathbf{g}$, which will inevitably change the deterministic cost, $Q( \mathbf{g})$, into a random cost, $Q( \mathbf{g,\mathbf{p}(\bm{\omega})})$. In this paper, we consider the randomness brought by the wind farms as a spatiotemporal random field, which is denoted by $\mathbf{p}_{i}(\bm{\omega},\mathbf{t})$ for the power generation of the $i$th wind farm. Here, the time $t \in \mathbb{T}$ and $\mathbb{T}$ is a finite integer set representing hours in a day, namely,   $\mathbb{T}=\{1,2,\dots,24\}$.
Let us take an example. Suppose that we conduct a day-ahead SED problem over 24 hours of a power system with three farms. Then, we have an input of three random fields, \{$\mathbf{p}_1(\bm{\omega},\mathbf{t})$, \{$\mathbf{p}_2(\bm{\omega},\mathbf{t})$, \{$\mathbf{p}_3(\bm{\omega},\mathbf{t})$\}, consisting of  $72$ random variables in total.  Our uncertainty quantification goal is to quantify the  statistical moments of $Q( \mathbf{g,\mathbf{p}(\bm{\omega})})$, such as the mean and variance for a day-ahead forecast.

\begin{remark}
Note that since we focus on the quantification of uncertainties in the SED problem instead of their modeling, the detailed description of ``$f$" as well as all the equality and inequality constraints of the SED topic directly follow from~\cite{safta2016efficient}. 
\end{remark}

\vspace{-0.15cm}
\section{Theoretical Background}
In this section, we briefly present the theory of the GPE and of the KLE to assist us in the SED problem. 
\vspace{-0.25cm}
\subsection{Gaussian Process Emulator}
\vspace{-0.1cm}
\subsubsection{Basic Theory} The GPE is known to be a powerful Bayesian-learning-based method based on a nonlinear regression problem~\cite{rasmussen2006gaussian,gelman2014bayesian}.  To describe this method, let us first denote the SED model by ${f(\cdot)}$ and its corresponding vector-valued input of $p$ dimensions by $\mathbf{x}$. Due to the randomness of  $\mathbf{x}$, we may observe $n$ samples as a  finite  collection of the model input as $\{\mathbf{x}_{1}, \mathbf{x}_{2}, \dots, \mathbf{x}_{n} \}$. 
Accordingly, its model output $f(\mathbf{x})$ also becomes random and has its corresponding $n$ realizations, denoted by $\{f(\mathbf{x}_{1}), f(\mathbf{x}_{2}), \dots, f(\mathbf{x}_{n})\}$.

If we assume that the model output is a realization of a Gaussian process, then the finite collection, $\{f(\mathbf{x}_{1}), f(\mathbf{x}_{2}), \dots, f(\mathbf{x}_{n})\}$, of the random variables,  $f(\mathbf{x})$, will follow a joint multivariate normal probability distribution, that is, we have
\begin{equation}
\label{eqn:2}
   \left[\begin{array}{c}{{\scriptstyle f\left(\mathbf{x}_{1}\right)}} \\ {\vdots} \\ {{\scriptstyle f\left(\mathbf{x}_{n}\right)}}\end{array}\right] \sim \scriptstyle{ \mathcal{N}} \left(\left[\begin{array}{c}{{\scriptstyle m\left(\mathbf{x}_{1}\right)}} \\ {\vdots} \\ {{\scriptstyle m\left(\mathbf{x}_{n}\right)}}\end{array}\right],\left[\begin{array}{ccc} {{\scriptstyle k\left(\mathbf{x}_{1}, \mathbf{x}_{1}\right)}} & {\cdots} & {{\scriptstyle k\left(\mathbf{x}_{1}, \mathbf{x}_{n}\right)}} \\ {\vdots} & {\ddots} & {\vdots} \\ {{\scriptstyle k\left(\mathbf{x}_{n}, \mathbf{x}_{1}\right)}} & {\cdots} & {{\scriptstyle k\left(\mathbf{x}_{n}, \mathbf{x}_{n}\right)}}\end{array}\right]\right).
\end{equation}
Here, $\bm{m}(\bm{\cdot})$ is the mean function and $\bm{k}(\bm{\cdot},\bm{\cdot})$ is a kernel function that represents the covariance function.  Let us further denote an $n \times p$ matrix, denoted by $\mathbf{X} = [\mathbf{x}_{1}, \mathbf{x}_{2}, \dots, \mathbf{x}_{n}]^{\transpose}$. Then, \eqref{eqn:2} is simplified into
\begin{equation}
\label{eqn:3}
\bm{f}\left(\mathbf{X}\right) | \mathbf{X} \sim \mathcal { N }\left(\bm{m}\left(\mathbf{X} \right), \bm{k}\left(\mathbf{X}, \mathbf{X}\right)\right),
\end{equation}
where $\bm{f}(\mathbf{X}) = (f(\mathbf{x}_{1}), f(\mathbf{x}_{2}), \dots, f(\mathbf{x}_{n}))^{\transpose}$
and $\bm{m}(\mathbf{X}) =(m(\mathbf{x}_{1}), m(\mathbf{x}_{2}), \dots, m(\mathbf{x}_{n}))^{\transpose}$.

Now, if an observation noise $\boldsymbol{\varepsilon}$ is added to $\bm{f}(\mathbf{X})$, we get 
\begin{equation}
\label{eqn:4}
    \mathbf{Y} = \bm{f}(\mathbf{X})+\boldsymbol{\varepsilon}.
\end{equation}
For independent, identically and normally distributed noise $\boldsymbol{\varepsilon} \sim \mathcal{N}(0, \sigma^{2}\mathbf{I}_{n})$
 (where $\mathbf{I}_{n}$ and $\sigma^{2}$ are an $n$-dimensional identity matrix and the variance, respectively), using normality property, we obtain 
\begin{equation}
\label{eqn:5}
\mathbf{Y} | \mathbf{X} \sim \mathcal { N }\left(\bm{m}\left(\mathbf{X}\right), \bm{k}\left(\mathbf{X}, \mathbf{X}\right) + \sigma^{2}\mathbf{I}_{n} \right).
\end{equation}
Note that $\boldsymbol{\varepsilon}$ is also called a ``nugget''. If $\sigma^{2} = 0$, then  $f(x)$ is observed without noise.  However, in practical implementation, the nugget is always added for the sake of numerical stability.

\vspace{-0.15cm}
\subsubsection{Bayesian Inference}
 Here, we present the way to use the abovementioned finite collection of $n$ samples, $\{\mathbf{Y}, \mathbf{X}\}$, to infer the unknown system output,  $\mathbf{y}(\mathbf{x})$, on the sample space of $\mathbf{x} \in \mathbb{R}^{p}$ in a Bayesian inference framework.  
 We assume that the readers have the basic knowledge of Bayesian inference. 
 
 Here, the finite collection of samples $\{\mathbf{Y}, \mathbf{X}\}$ provides us with the observations. To infer a Bayesian posterior distribution of the unknown system output  $\mathbf{y}(\mathbf{x})$, we must assume a Bayesian prior distribution of $\mathbf{y}(\mathbf{x})$, expressed as
\begin{equation}
\label{eqn:6}
\mathbf{\mathbf{y}(\mathbf{x})} | \mathbf{x} \sim \mathcal { N }\left(\bm{m}\left(\mathbf{x}\right), \bm{k}\left(\mathbf{x}, \mathbf{x}\right) + \sigma^{2}\mathbf{I}_{n_{x}} \right).
\end{equation}

\begin{figure}[t]
\centering
\subfloat[{\label{fig:1a}}]
{\includegraphics[width=0.24\textwidth]{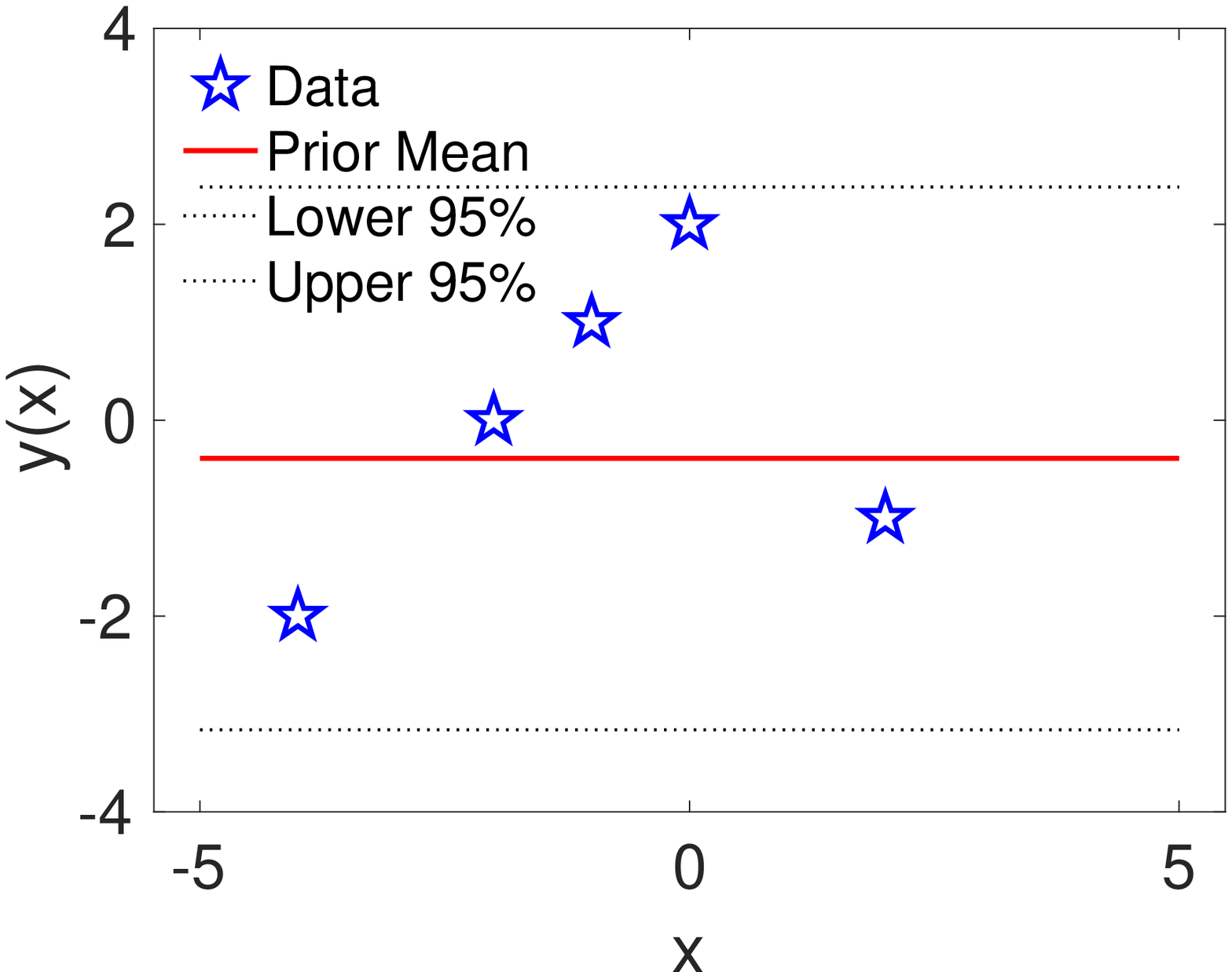}}\hfill
\subfloat[\label{fig:1b}]{\includegraphics[width=0.24\textwidth]{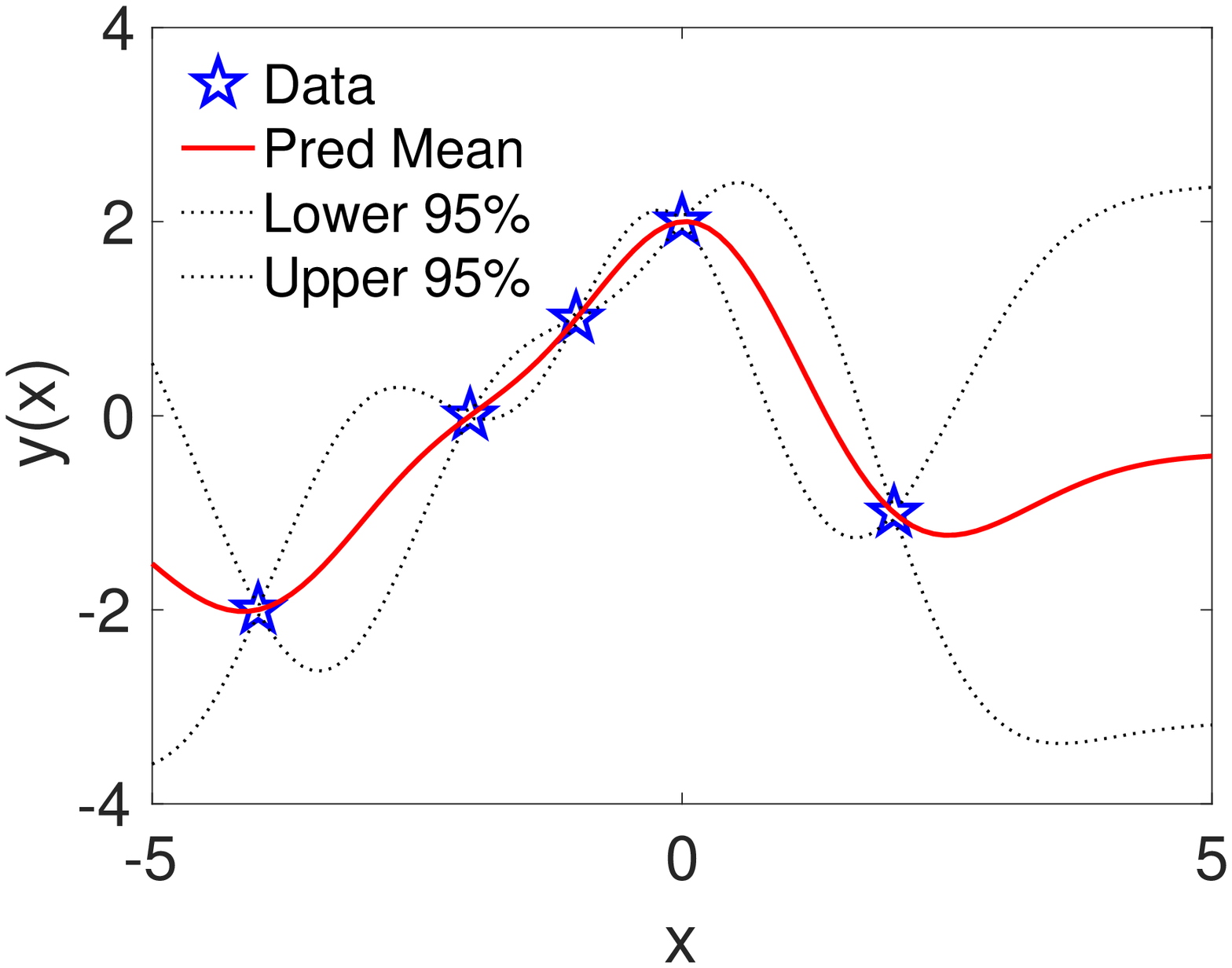}}
\vspace{-0.1cm}
\caption{(a) The prior and (b) updated posterior means as well as 95\% confidence intervals (CIs) of a GP with constant mean function and squared exponential kernel. The posterior captures much more information of data than the prior does.} \label{fig:1}
\end{figure}

Then, we can formulate the joint distribution of $\mathbf{Y}$ and $\mathbf{y}(\mathbf{x})$ using \eqref{eqn:5} and \eqref{eqn:6} as
\begin{equation}
\label{eqn:7}
\left[\begin{array}{c}{{\mathbf{Y}}}  \\ { \mathbf{y}(\mathbf{x})}\end{array}\right] \sim { \mathcal{N} \left(\left[\begin{array}{c}{\bm{m}\left(\mathbf{X}\right)}  \\  \bm{m}\left(\mathbf{x}\right)\end{array}\right],\left[\begin{array}{cc} { \mathbf{K}_{11} }  & {\mathbf{K}_{12}} \\   {\mathbf{K}_{21}} & {\mathbf{K}_{22}} \end{array}\right]\right),
}
\end{equation}
where $ \mathbf{K}_{11} = \bm{k}\left(\mathbf{X}, \mathbf{X}\right) + \sigma^{2}\mathbf{I}_{n}$, $\mathbf{K}_{12} = \bm{k}\left(\mathbf{X}, \mathbf{x}\right)$, $\mathbf{K}_{21} = \bm{k}\left(\mathbf{x}, \mathbf{X}\right)$ and $\mathbf{K}_{22} = \bm{k}\left(\mathbf{x}, \mathbf{x}\right) + \sigma^{2}\mathbf{I}_{n_{x}}$. 
Now, we can infer $\mathbf{y}(\mathbf{x})$ based on previous observations $\left(\mathbf{Y}, \mathbf{X}\right)$. 
Using  the  rules of the conditional Gaussian distribution (a.k.a. Gaussian conditioning or statistical linearization) \cite{eaton1983multivariate},  the Bayesian posterior distribution of the system output  $\mathbf{y}(\mathbf{x})$ conditioned upon the observations $\left(\mathbf{Y}, \mathbf{X}\right)$ follows a Gaussian distribution given by
\begin{equation}
\label{eqn:8}
\mathbf{y}(\mathbf{x}) | \mathbf{x}, \mathbf{Y}, \mathbf{X} \sim \mathcal { N }\left( \bm{\mu} \left(\mathbf{x} \right), \bm{\Sigma} \left( \mathbf{x} \right) \right),
\end{equation}
where 
\begin{equation}
\label{gpemean}
    \bm{\mu} \left(\mathbf{x} \right) = \bm{m}(\mathbf{x}) + \mathbf{K}_{21}\mathbf{K}_{11}^{-1} (\mathbf{Y} - \bm{m}(\mathbf{X})), 
\end{equation}
\begin{equation}
\label{gpecov}
    \bm{\Sigma} \left( \mathbf{x} \right) =  \mathbf{K}_{22} - \mathbf{K}_{21} \mathbf{K}_{11}^{-1} \mathbf{K}_{12}.
\end{equation}
To this point, the form of the GPE has been derived. Now, on one hand, we may directly use  \eqref{gpemean} as a surrogate model (a.k.a. the response surface or reduced-order model) to capture very closely the behavior of the complicated, original simulation model of a power system while being computationally inexpensive to evaluate. On the other hand, we may use \eqref{gpecov} to quantify the uncertainty of the surrogate itself. In this paper, we only need to use \eqref{gpemean} as a surrogate model.
 Fig. \ref{fig:1} shows a simple example of how five 1-dimensional data points update a Gaussian process prior to a posterior.
 

\subsubsection{Mean and Covariance Functions}
To further define the GPE, we need to select the forms of the mean function $\bm{m}(\bm{\cdot})$
and the covariance function represented via the kernel $\bm{k}(\bm{\cdot},\bm{\cdot})$. 

The mean function models the prior belief about the existence of a systematic trend expressed as
\begin{equation}
\label{trend}
    \bm{m}(\mathbf{x}, \boldsymbol{\beta}) = \mathbf{H}(\mathbf{x}) \boldsymbol{\beta}.
\end{equation}
Here, $\mathbf{H}(\mathbf{x})$ can be any set of basis functions. For example, let $ \mathbf{x}_{i} = (x_{i1},\dots,x_{ip})$ indicate the $i$th sample, $i=1,2,\dots,n$ and $x_{ik}$ represents its $k$th element, $k=1,2,\dots, p$. For instance, $\mathbf{H}(\mathbf{x}_{i}) = 1$ is a constant basis; $\mathbf{H}(\mathbf{x}_{i}) =(1, x_{i 1},\dots, x_{i p})$ is a linear basis; $\mathbf{H}(\mathbf{x}_{i}) = (1, x_{i 1}, \dots, x_{i p}, x_{i 1}^{2},\dots, x_{i p}^{2})$ is a pure quadratic basis; and $\boldsymbol{\beta}$ is a vector of hyperparameters. 

Since the covariance function is represented by a kernel function, choosing the latter is a must. Table \ref{t: 6} provides several popular covariance kernels. 
\begin{table}[ht]
\renewcommand{\arraystretch}{1.3}
\caption{ Commonly Used Covariance Kernels for Gaussian Process}
\label{t: 6}
\centering
\begin{tabular}{rll}
\toprule
\hline

$k_{\text{SE}}\left(\mathbf{x}_{i}, \mathbf{x}_{j}\right)$ & $\tau^{2} \exp \left(- \sum\limits_{k=1}^{p} \frac{r_{k}^{2}}{2 \ell_{k}^{2}}\right) $ \\
$k_{\text{E}}\left(\mathbf{x}_{i}, \mathbf{x}_{j}\right)$ & $\tau^{2} \exp \left(- \sum\limits_{k=1}^{p} \frac{\left| r_{k} \right|}{ \ell_{k}}\right)$ \\ 
$k_{\text{RQ}}\left(\mathbf{x}_{i}, \mathbf{x}_{j}\right)$& $\tau^{2}\left(1+ \sum\limits_{k=1}^{p} \frac{r_{k}^{2}}{2 \alpha \ell_{k}^{2}}\right)^{ - \alpha}$ \\ 
$k_{3/2}\left(\mathbf{x}_{i}, \mathbf{x}_{j} \right)$& $\tau^{2}\left(1+ \sum\limits_{k=1}^{p} \frac{\sqrt{3} r_{k}}{\ell_{k}}\right) \exp \left(-\sum\limits_{k=1}^{p} \frac{\sqrt{3} r_{k}}{\ell_{k}}\right)$ \\ 
&  $(r_{k} = \left|x_{ik}-x_{jk}\right|)$ \\
\bottomrule
\end{tabular}
\vspace{1mm}

Abbrv.: square exponential (SE), exponential (E), rational quadratic (RQ), and Martin 3/2 ($3/2$) kernels. 

\end{table}

As for the parameters of a kernel function, they are defined as follows: $\tau$ and $\ell_{k}$ are the hyperparameters defined in the positive real line; $\sigma^{2}$ and $\ell_{k}$ correspond to the order of magnitude and the speed of variation in the $k$th input dimension, respectively. Let $\boldsymbol{\theta} = (\tau, \ell_{1}, \dots, \ell_{p} )$ contains the hyperparameters of the covariance function, i.e.,
\begin{equation}
\label{kerneltype}
k\left(\mathbf{x}_{i}, \mathbf{x}_{j} | \boldsymbol{\theta} \right)  = \Cov(\mathbf{x}_{i}, \mathbf{x}_{j} | \boldsymbol{\theta}).
\end{equation}

Until now, the model structure of the GPE has been fully defined. For simplicity, we write $\boldsymbol{\eta} = (\sigma^{2}, \boldsymbol{\beta}, \boldsymbol{\theta})$ to represent all the hyperparameters in the GPE model. 

\subsubsection{Hyperparameter Estimation} Optimizing a GPE model is equivalent to estimating $\boldsymbol{\eta}$ given the data $\left(\mathbf{Y}, \mathbf{X}\right)$. Although different methods exist for estimating the hyperparameters $\bm{\eta}$ \cite{gelman2014bayesian}, we choose to adopt the Gaussian maximum likelihood estimator (MLE) since it meets our demand and is straightforward to compute.

First, to indicate the hyperparameters, let us rewrite \eqref{eqn:5} as
\begin{equation}
\label{reform}
\mathbf{Y} | \mathbf{X},\boldsymbol{\eta} \sim \mathcal { N }\left(\bm{m}\left(\mathbf{X}\right), \bm{k}\left(\mathbf{X}, \mathbf{X}\right) + \sigma^{2}\mathbf{I}_{n} \right). 
\end{equation}
Then, using MLE, we obtain
\begin{equation}
\label{eqn:11}
    \widehat{\boldsymbol{\eta}} = \left(\widehat{\boldsymbol{\beta}}, \widehat{\boldsymbol{\theta}}, \widehat{\sigma}^{2} \right)=\underset{\boldsymbol{\beta}, \boldsymbol{\theta}, \sigma^{2}}{\arg \max } \log P\left(\mathbf{Y} | \mathbf{X}, \boldsymbol{\beta}, \boldsymbol{\theta}, \sigma^{2}\right).
\end{equation}
Using \eqref{trend}--\eqref{reform} and simplifying $\mathbf{H}(\mathbf{x})$ into $\mathbf{H}$, the marginal log-likelihood can be expressed as
\begin{equation}
\label{eqn:12}
\begin{aligned} & \log P\left(\mathbf{Y} | \mathbf{X}, \boldsymbol{\beta}, \boldsymbol{\theta}, \sigma^{2} \right) \\
=&-\frac{1}{2}(\mathbf{Y}-\mathbf{H} \boldsymbol{\beta})^{\transpose}\left[\bm{k}(\mathbf{X}, \mathbf{X} | \boldsymbol{\theta})+\sigma^{2} \mathbf{I}_{n}\right]^{-1}(\mathbf{Y}-\mathbf{H} \boldsymbol{\beta}) \\ &-\frac{n}{2} \log 2 \pi-\frac{1}{2} \log \left|\bm{k}(\mathbf{X}, \mathbf{X} | \boldsymbol{\theta})+\sigma^{2} \mathbf{I}_{n}\right|, 
\end{aligned}
\end{equation}
which implies that the MLE of $\boldsymbol{\beta}$ conditioned upon $\boldsymbol{\theta}$ and $\sigma^{2}$ is a weighted least-squares estimate given by 
\begin{equation}
\label{eqn:13}
    {\scriptstyle \hat{\boldsymbol{\beta}}\left(\boldsymbol{\theta}, \sigma^{2}\right)=} {\scriptstyle\left[\mathbf{H}^{\transpose}\left[\bm{k}(\mathbf{X}, \mathbf{X} | \boldsymbol{\theta})+\sigma^{2} \mathbf{I}_{n}\right]^{-1} \mathbf{H}\right]^{-1} \mathbf{H}^{\transpose}\left[\bm{k}(\mathbf{X}, \mathbf{X} | \boldsymbol{\theta})+\sigma^{2} \mathbf{I}_{n}\right]^{-1} \mathbf{Y}}.
\end{equation}
Plugging \eqref{eqn:13} into \eqref{eqn:12}, we get the $\boldsymbol{\beta}$-profile likelihood $\log P\left(\mathbf{Y} | \mathbf{X}, \hat{\boldsymbol{\beta}}\left(\boldsymbol{\theta}, \sigma^{2}\right), \boldsymbol{\theta}, \sigma^{2} \right)$. Then, \eqref{eqn:11} is rewritten as
\begin{equation}
\label{eqn:14}
    \left(\widehat{\boldsymbol{\theta}}, \widehat{\sigma}^{2}\right) =\underset{\boldsymbol{\theta}, \sigma^{2}}{\arg \max } \log P\left(\mathbf{Y} | \mathbf{X}, \hat{\boldsymbol{\beta}}\left(\boldsymbol{\theta}, \sigma^{2}\right), \boldsymbol{\theta}, \sigma^{2}\right), 
\end{equation}
where $\widehat{\boldsymbol{\beta}} =  \hat{\boldsymbol{\beta}}\left(\widehat{\boldsymbol{\theta}}, \widehat{\sigma}^{2} \right)$. The next goal is to find $\widehat{\boldsymbol{\eta}}$ from \eqref{eqn:12}--\eqref{eqn:14}. Since $\widehat{\boldsymbol{\beta}}$ can be straightforwardly obtained from $\left(\widehat{\boldsymbol{\theta}}, \widehat{\sigma}^{2}\right)$, one only needs to find $\left(\widehat{\boldsymbol{\theta}}, \widehat{\sigma}^{2}\right)$ by maximizing the $\boldsymbol{\beta}$-profile likelihood over $\left(\boldsymbol{\theta}, \sigma^{2} \right)$. Here, we utilize  a gradient-based optimizer to achieve this optimization. To overcome the presence of local optima in the objective function, we initialize $\sigma^{2}$ somewhere close to 0 because the global optimum is in this vicinity. Once $\widehat{\boldsymbol{\eta}}$ is obtained, the GPE model is fully constructed. 

\vspace{-0.1cm}
\subsubsection{Sampling Strategy}
In order to obtain the observation sets contained in $\left(\mathbf{Y}, \mathbf{X}\right)$, we need some samples that satisfy the system function $\bm{f}(\mathbf{X})$, i.e., the  SED model. Here, a popular choice to generate these samples is through the Latin hypercube sampling ~\cite{santner2003design}. Unlike the Monte Carlo sampling, which generates a set of independent and identically distributed samples from the target probability distributions, the Latin hypercube sampling generates near-random samples that follow a standard uniform distribution $\mathcal{U}(0, 1)$ based on an equal-interval segmentation.  For a nonuniform distribution, the inverse transformation of the cumulative distribution function is applied to map the uniformly distributed samples into the targeted distribution\cite{devroye1986sample}. 

\vspace{-0.3cm}
\subsection{Karhunen-Lo\`eve Expansions}
As it is mentioned in Section II, the dimension for the random fields representing wind-farm generation may be so high that the GPE cannot be constructed efficiently. Therefore, facing the challenge raised by a high-dimensional raw data, an efficient dimension reduction becomes a prerequisite.  
\vspace{-0.2cm}
\subsubsection{Spectral Decomposition and Truncation}
Here, we  use the KLE to project the high-dimensional samples  into  low-dimensional latent  variables.  Let us consider a bounded domain $D \subseteq \mathbb{R}$ and a sample space $\Omega$, and let $X(t)$ be a zero-mean stochastic process with $t \in D$ where $X : D \times \Omega \rightarrow \mathbb{R}$. Each $X(t)$ is a random variable indexed by $t$. Let us assume that $X(t)$ for any time has a finite variance and let us define the covariance function $C$ as $C(t, s) = \Cov(X(t), X(s)), \forall t, s \in D$. Since $C$ is positive definite, its spectral decomposition is obtained as 
\begin{equation}
\label{eqn:16}
    C(t, s)=\sum_{l=1}^{\infty} \lambda_{i} u_{i}(t) u_{i}(s).
\end{equation}
Here, $\lambda_{i}$ denotes the $i$th eigenvalue  and $u_{i}$ denotes the $i$th orthonormal eigenfunction  of $C$.  Then, we put $X(t)$ into a KLE framework as follows:
\begin{equation}
\label{eqn:17}
    X(t)= \sum_{i=1}^{\infty} \sqrt{\lambda_{i}} u_{i}(t) \xi_{i},
\end{equation}
where \{$\xi_{i}, i= 1, 2, ...$\} are mutually uncorrelated univariate random variables with zero mean and unit variance. Here, we use an empirical covariance matrix, $C$, calculated from the data. By applying the inner product of $u_{i}(t)$ to both sides of \eqref{eqn:17} and making $u_{i}(t)$ orthonormal, we obtain 
\begin{equation}
\label{eqn:18}
      \sqrt{\lambda_{i}} \xi_{i} = \langle X(t), u_{i}(t) \rangle \quad \forall i. 
\end{equation}
Till now, the KLE maps $X(t)$ to latent variables $\xi_{i}$ by projecting $X(t)$ onto $u_{i}(t)$. The inverse mapping can be achieved via \eqref{eqn:17} as well. Note that both transformations are linear. In practice, $X(t)$ is replaced by a finite summation with $p$ elements, yielding  
\begin{equation}
\label{eqn:19}
    X(t)\approx \hat{X}(t) = \sum_{i=1}^{p} \sqrt{\lambda_{i}} u_{i}(t) \xi_{i}.
\end{equation}
\subsubsection{Dimension Reduction}
Here, we present the dimension reduction from the variance point of view.  Consider the total variance of $X(t)$ over $D$. Using the  orthonormality property of $u_{i}(t)$, we have
\begin{equation}
\label{varx}
\int _{D} \Var[ X(t) ] \d t = \sum_{i=1}^{\infty} \lambda_{i} \\.
\end{equation}
A finite series is used instead such that most of the variance is retained after truncation, yielding
\begin{equation}
\label{pvarx}
\int _{D} \Var[ X(t) ] \d t 	\approx  \int _{D} \Var[ \hat{X}(t) ] \d t =  \sum_{i=1}^{p} \lambda_{i}. 
\end{equation}
Specifically, we choose the first $p$ largest eigenvalues as $(\lambda_{1},\dots,\lambda_{p})$, with the corresponding eigenfunctions as $(u_{1},\dots,u_{p})$ such that the obtained $\boldsymbol{\xi}$ contain over 95\% of the total variance calculated by $ \left(\sum_{i=1}^{p} \lambda_{i}\right) /\left(\sum_{i=1}^{\infty} \lambda_{i}\right) $.
Since the calculated $p$ is smaller than the original dimension of the raw data sequence, the KLE has mapped the high-dimensional correlated $X$ to the low-dimensional uncorrelated $\boldsymbol{\xi} = (\xi_{1}, \dots, \xi_{p})$. It is worth pointing out that the truncated $p$ dimensions will serve as the input for the SED problem. 

\vspace{-0.3cm}
\section{Proposed Method}
Using the theory explained above, we propose a GPE-based method to solve the SED problem. We first obtain $\bm{\xi}$ of the $p$-truncated KLE for the SED input. Then, to avoid the normality assumption, we estimate the closed-form solution of the joint probability density function (pdf) of $\bm{\xi}$ via a kernel density estimation \cite{silverman1998}. Later, using that density estimation, the input samples are regenerated. Finally, with the training samples selected from the Latin hypercube sampling, the GPE surrogate is constructed to propagate the uncertainty from the input samples to the output. The details are described in Algorithm 1.
Note that since the input of the SED is spatiotemporally correlated, our approach is naturally compatible with the modeling of a spatiotemporal structure. The local $\bm{\xi}$ is calculated by applying the KLE individually to each location. After estimating the joint pdfs of $\bm{\xi}$ at all locations, samples can be drawn at every location for each time step.

\begin{algorithm}[!htbp]
\label{MF-PCE-AIS}
\caption{The Proposed GPE-based SED Method }

\begin{algorithmic} [1]
\State  Read the historical raw data of the wind farms;

\State Apply the KLE on the preprocessed data to obtain $p$-dimensional truncated data;

 \State Estimate the closed-form solution for the joint pdf of $\bm{\xi}$; 
 
\State Regenerate a large number of samples as the model input through the pdf obtained from the kernel density estimation; 

\State Generate $p$-dimensional $n$ training samples $\mathbf{X}$ via the Latin hypercube sampling method;

\State Obtain realizations $\mathbf{Y}$ by evaluating $\mathbf{X}$ through the SED model;

\State Estimate the hyperparameters $\boldsymbol{\eta}$ given the  training set $\left(\mathbf{Y}, \mathbf{X}\right)$ for the GPE-based surrogate model;

\State Propagate a large number of samples of the model input through the GPE-based surrogate models to obtain the SED system output realizations; 

\State Calculate the  statistical moments of the SED output $Q(g, \mathbf{p}(\bm{\xi}))$.

\end{algorithmic}
\end{algorithm}

\vspace{-0.3cm}
\section{Case Studies}
We test our method on the IEEE 118-bus system\cite{118system} with the \textsc{Matpower} package using the MATLAB\textsuperscript{\textregistered} R$2019$a version and the NREL's Western Wind Data Set \cite{NREL}. In the experiment, we pick one farm in Livermore, CA ($\#\mathrm{LV}$) and two farms in Seattle, WA ($\#\mathrm{SE1}, \#\mathrm{SE2}$). For each farm, three turbines are extracted from the dataset: $\#9247, \#9248, \#9249$ for $\#\mathrm{LV}$; $\#28914, \#28928, \#28959$ for $\#\mathrm{SE1}$; and $\#29138, \#29153, \#29154$ for $\#\mathrm{SE2}$. These three wind farms are added at Buses 16, 58, and 78, respectively. 

Assuming that the turbines in the same farm have the same wind speed and wind power all the time, we calculate for each time stamp the averaged wind speed and wind power over three selected wind turbines and treat it as the prevailing wind speed and wind power of that farm. For each farm, we take hourly averages on the common wind speed and wind power to obtain the daily wind speed $W$ and wind power $P$ for each day in January between 2004 and 2006, which leads to a total of 93 data points. The relationship between $P$ and $W$ is modeled by a decision tree regression model. The KLE is used to represent the randomness of $W$. To preserve 95\% of the total variance of $W$, the first 8, 4, and 5 KLE modes are kept for $\#\mathrm{LV}, \#\mathrm{SE1}, \text{and} \ \#\mathrm{SE2}$, respectively.

To test the spatial dependency of $W$ among the three farms, we calculate distance correlation factors~\cite{szekely2007measuring} between $\boldsymbol{\xi}$ on different farms. The results are shown in Appendix, where a way to deal with the dependency between $\#\mathrm{SE1}$ and $\#\mathrm{SE2}$ is also provided.
\begin{figure}[htbp]
\centering
\vspace{-0.15cm}
\subfloat[\label{fig:2a}]{\includegraphics[width=0.24\textwidth]{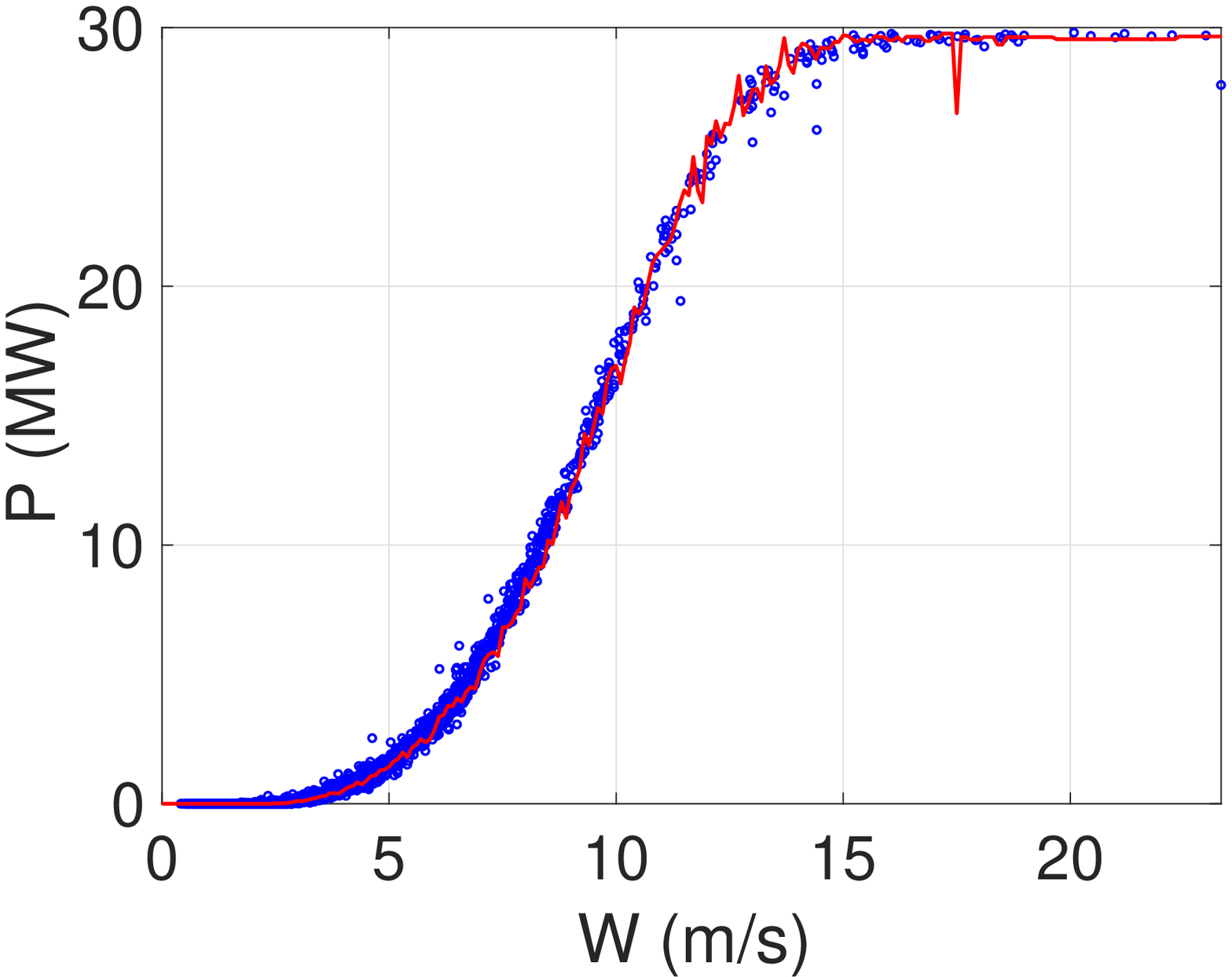}}
\hfill
\subfloat[\label{fig:2b}]{\includegraphics[width=0.24\textwidth]{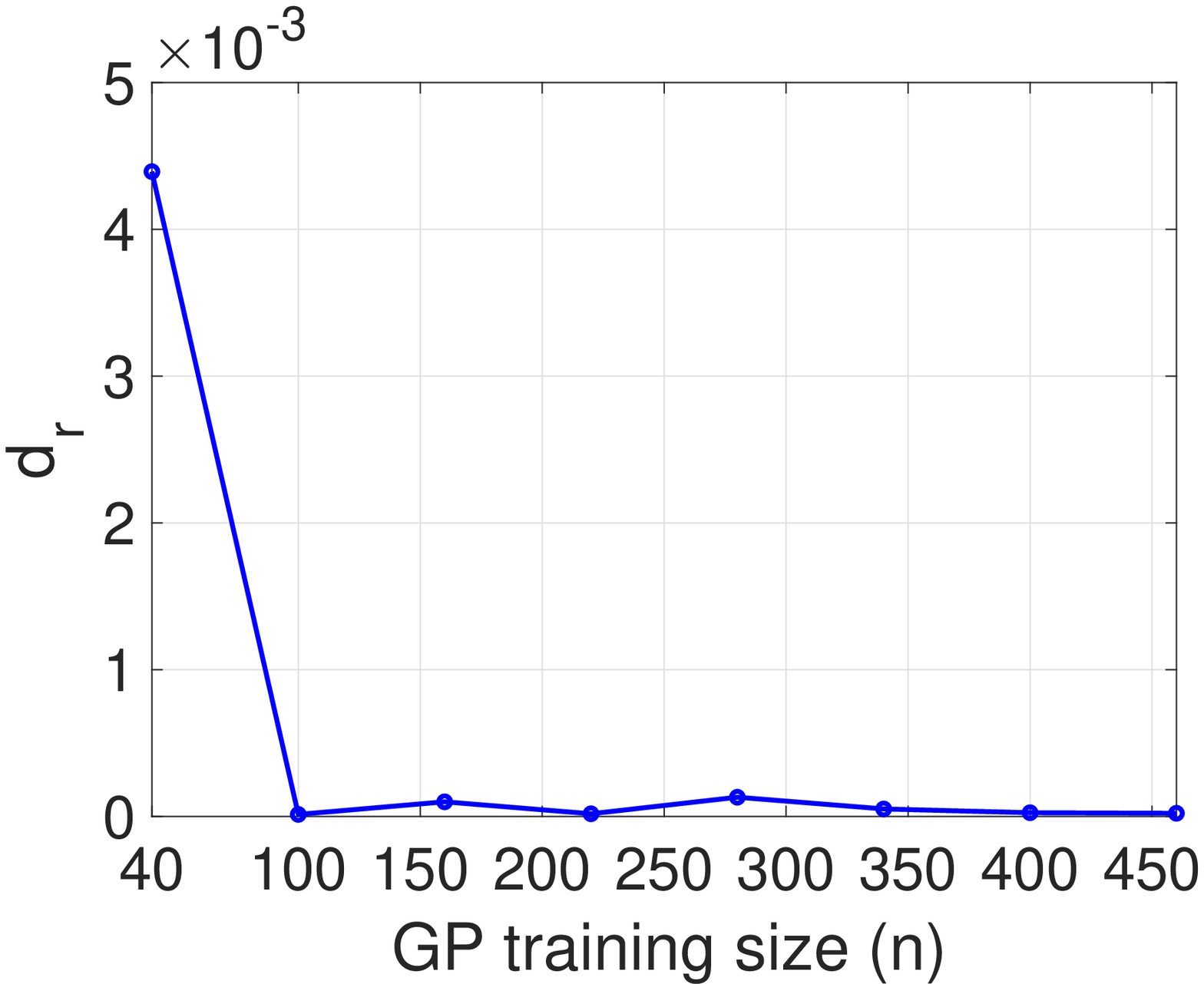}}
\hfill
\vspace{-0.1cm}
\caption{(a) Wind power $P$ versus wind speed $W$ at Farm $\#\mathrm{LV}$ in January (\textcolor{blue}{Blue} points denote real data points; whereas, \textcolor{red}{red} curve represents predictions by decision tree regression.). (b) The relative difference between $\E[Q_{\text{GP}}]$ and $\E[Q_{\text{MC}}]$ varying by $n$. In order to ensure that the MC converges, 8,000 realizations are used for $Q_{\text{MC}}$.} \label{fig:2}
\end{figure}


The GPE surrogate with a pure quadratic mean function and a squared exponential kernel is constructed for the SED test system. Let $Q_{\text{GP}}$ denote the estimation of the minimum production cost $Q$ from the GPE surrogate. $Q_{\text{MC}}$ is calculated by direct Monte Carlo simulations performed on the test system. One of the most important results obtained from the SED is the expected minimum cost $\E[Q]$. We define the relative difference between $\E[Q_{\text{GP}}]$ and $\E[Q_{\text{MC}}]$ as
$ d_r = \frac{\lvert  \E[Q_{\text{GP}}] - \E[Q_{\text{MC}}] \rvert }{\E[Q_{\text{MC}}]}\cdot\E[Q_{\text{MC}}]$ is a fixed baseline while $\E[Q_{\text{GP}}]$ varies with the GPE training size $n$. The smaller the $d_r$ is, the better the GPE surrogate fits its target. 

\begin{table}[!htbp]
\renewcommand{\arraystretch}{1.2}
\caption{Predictive Inferences (Mean, 95\% CI and Standard Deviation) of Minimum Production Cost (GP Training Size $n=100$.)}
\label{t:4}
\centering
\begin{tabular}{lcccc}
\toprule
 & Mean $( \times 10^{6} )$ & 95\% CI $ ( \times 10^{6})$ 
 & Std. Dev. $( \times 10^{4})$\\
\hline
$Q_{\text{MC}}$ & $2.955$ & $\left(2.874, 3.018\right)$ & $3.908$ \\ 
$Q_{\text{GP}}$& $2.956$ & $\left(2.869, 3.020\right)$ & $4.078$\\ 
\bottomrule
\end{tabular}

\end{table}

Figure~\ref{fig:2}\subref{fig:2b} depicts the trace plot of $d_r$ over $n$. Notice that the approach achieves less than $10^{-3} d_r$ when $n=100$. Numerical inferences are listed in Table \ref{t:4}. The GPE surrogate, trained with 100 Latin hypercube simulations, successfully calculates the SED values for the tested system under 8,000 scenarios.


\begin{table}[ht]
\renewcommand{\arraystretch}{1.2}
\caption{ Results from Replicating Empirical Minimum Production Cost (GP Training Size $n=100$.)}
\label{t:5}
\centering
\begin{tabular}{lcccc}
\toprule
 & Mean $ (\times 10^{6})$ & 95\% CI $( \times 10^{6})$ 
 & Std. Dev. $( \times 10^{4})$\\
\hline
$Q_{\text{data}}$ & $2.943$ & $\left( 2.849, 3.017 \right)$ & $4.996$ \\
$Q^{r}_{\text{GP}}$& $2.949$  & $\left( 2.858 , 3.025 \right)$ & $5.108$ \\ 
\bottomrule
\end{tabular}
\end{table}

Ideally, a well-performing surrogate is also expected to reasonably replicate the empirical cost $Q_{\text{data}}$ for 93 data points. Correspondingly, we have the replications $Q_{\text{GP}}$ from the GPE-based  surrogate estimation. The results in Table \ref{t:5} indicate that the trained GPE surrogate is able to reproduce the empirical $Q$ of the test system. 

\vspace{-.2cm}
\section{Conclusions and Future Work}
In this paper, we propose a GPE-based framework in quantifying uncertainty for the SED problem. The proposed framework utilizes the KLE to conduct an effective dimension reduction, which further accelerates the nonparametric GPE in the propagation of uncertainties. The simulation results on the modified IEEE 118-bus system show that the proposed method is significantly more computationally efficient than the traditional Monte Carlo method while achieving the desired simulation accuracy.

\vspace{-.2cm}
\appendix[Modeling Spatial Correlations of Wind Speeds between Wind Farms in Seattle]

\begin{table}[h]
\renewcommand{\arraystretch}{1.2}
\caption{Distance Correlations for the First Three of $\bm{\xi}$ between Wind Farms}
\label{t:3}
\centering
\begin{tabular}{lcccc}
\toprule
& $\#\mathrm{LV}$ -- $\#\mathrm{SE1}$ & $\#\mathrm{LV}$ -- $\#\mathrm{SE2}$ & $\#\mathrm{SE1}$ -- $\#\mathrm{SE2}$\\
\hline
$\xi_{1}$ & $0.141$ & $0.123$ & $0.592$ \\
$\xi_{2}$ & $0.221$ & $0.166$ & $0.289$ \\
$\xi_{3}$ & $0.131$ & $0.231$ & $0.319$ \\ 
\bottomrule
\end{tabular}
\end{table}
Table \ref{t:3} shows that $\xi_{1}$ between $\# \mathrm{SE1}$ and $\#\mathrm{SE2}$ have a relatively high distance correlation. One may still treat them independently since the value is not too close to $1$. In addition, we provide a way to handle the dependency as described below. 

 Considering that the Pearson correlation of $\xi_{1}$ between $\#\mathrm{SE1}$ and $\#\mathrm{SE2}$ is calculated to be 0.618, it is proper to assume that they are linearly dependent, which can be modeled using a linear regression. If we use $\xi_{11}$ and $\xi_{12}$ to distinguish $\xi_{i}$ in $\#\mathrm{SE1}$ and $\#\mathrm{SE2}$, the joint samples of $\xi_{11}, \xi_{12}$ can be obtained in two steps: \begin{enumerate}
    \item Sample $\xi_{11}$ from its density function;
    
    \item Sample $\xi_{12} \sim \mathcal{N} (\hat{\beta_{0}} + \xi_{11}\hat{\beta_{1}}, \hat{\sigma}^{2})$, where $\hat{\beta_{0}}, \hat{\beta_{1}}, \hat{\sigma}^{2}$ are the estimated intercept, slope, and mean squared error from the results of the linear regression between $\xi_{12}$ and $\xi_{11}$, respectively.
\end{enumerate}

\vspace{-.2cm}
\section*{Acknowledgments}
This work was supported, in part, by the United States Department of Energy Office of Electricity Advanced Grid Modeling Program and performed under the auspices of the U.S. Department of Energy by Lawrence Livermore National Laboratory under Contract DE-AC52-07NA27344, and by the U.S. National Science Foundation under EPAS Grant 1917308. Document released as LLNL-CONF-788518.

\ifCLASSOPTIONcaptionsoff
  \newpage
\fi

\vspace{-.1cm}
\bibliographystyle{IEEEtran}
\bibliography{bibliography.bib}

%




\end{document}